\documentclass[reprint,prl,showpacs]{revtex4-1}
\usepackage{amssymb}
\usepackage{graphicx}
\usepackage{amsbsy}

\newcommand{\teff}{\tau_{\rm{eff}}}

\begin{document}

\title{Spin Resonance and dc Current Generation in a Quantum Wire}
\author{Ar. Abanov$^{a}$, V. L. Pokrovsky$^{a,b}$, W. M. Saslow$^{a}$, and
P. Zhou$^{a}$}
\affiliation{$^a$Department of Physics and Astronomy, TexasA\&M University, College
Station, TX 77843-4242\\
$^b$ Landau Institute for Theoretical Physics, Chernogolovka, Moscow Distr.
142432, Russia}

\begin{abstract}
We show that in a quantum wire the spin-orbit interaction leads to a
narrow spin resonance at low temperatures, even in the absence of an
external magnetic field. A relatively weak dc magnetic field of a
definite direction strongly increases the resonance absorption.
Linearly polarized resonance radiation produces dynamic
magnetization as well as electric and spin currents. The effect
strongly depends on the external magnetic field.
\end{abstract}

\pacs{73.21.Hb, 76.20.+q, 71.70.Ej}
\maketitle

\noindent \textbf{Introduction.} In recent years it has become
possible to engineer nanodevices using materials with predesigned 
properties.\cite{Wu2004,*Lauhon2004,nanowires,*nanowires1,*nanowires2,*nanowires3,*nanowires4}
This development revitalized an interest in electron interactions in
nanowires and led to discoveries of many fascinating phenomena. One
such interaction is the spin-orbit (SO) interaction of the
conduction electrons with the lattice. Even comparatively weak SO
interaction changes the symmetry of electronic system and leads to
numerous novel effects.

This Letter considers Electron Spin Resonance (ESR) in 1D nanowires
with a SO interaction.\cite{dressel,rashba,*BychkovRashba} The
standard picture of ESR in metals is as follows. In 3D an external
magnetic field $\mathbf{B}$ gives distinct 
up-spin and down-spin Fermi-surfaces, with
 Zeeman energy  the same for all electrons. An applied ac-field (considered to be almost uniform) induces resonant transitions between states with the same momentum (in the spherical shell between two Fermi-surfaces) and opposite directions of spin. However, a sufficiently strong SO interaction smears out this ESR
resonance.
The direction of the SO induced Zeeman ``internal'' magnetic field
$\mathbf{B}_{\rm SO}$ which acts on an electron depends on
the electron's momentum.

This Letter exploits the fact that this anisotropic broadening
is strongly reduced in         
a quantum wire where the direction of $\mathbf{B}_{\rm
SO}$ is the same for all $\mathbf{p}$. Thus ESR is 
narrow at low temperatures. This picture is the basis of our main results: The
relative strength of the Dresselhaus and Rashba interactions sets a
specific direction of $\mathbf{B}_{\rm SO}$ for a wire. Even a weak
dc magnetic field perpendicular to this direction increases the
resonance absorption by orders of magnitude, while the resonance
frequency $\omega_{r}$ changes only slightly. The component of
external magnetic field parallel to $\mathbf{B}_{\rm SO}$ separates
the resonances for left and right movers. Linearly
polarized resonance radiation then produces a net magnetization and
dc electric and spin currents. The magnitude of the effect is
controlled by the external magnetic field $\mathbf{B}$. For 1D
nanowires the geometrical constraints, together with quantization of
the transverse motion, strongly suppresses the most effective
Dyakonov-Perel mechanism of spin
relaxation,\cite{dyakonov-perel,*dyakonov-perel1} thus stabilizing
the resulting resonance induced non-equilibrium state.

The wire
can be formed by the growth process,\cite{nanowires,*nanowires1,*nanowires2,*nanowires3,*nanowires4}
 or from a semiconducting film or heterojunction by a proper
configuration of the gate electrodes.\cite{1d-from-2d} In the latter
case the 
substrate must violate reflection symmetry.

\noindent \textbf{Electronic spectrum and eigenstates.} Consider the type III-V
semiconductors GaAs and InGaAs.\footnote{We consider only their electron bands, to avoid complications associated with degeneracy of the hole band.} The electron
density is assumed to be sufficiently large and the temperature sufficiently
low to ensure a degenerate Fermi gas.
Electron-electron Coulomb interactions, i.e. Luttinger liquid effects in a 1D electron,
\cite{zulicke,*zulicke1} will be neglected.
We also assume that the wire is narrow enough to exclude multiple
channels.

In 1D the most general form of the SO interaction, including
both Rashba and Dresselhaus terms, is $H_{\rm SO}=\left( \alpha \sigma _{x}+\beta \sigma _{y}\right) p$, where $p$ is the 1D momentum along the wire direction $x$,\footnote{$p_{y}$ and $p_{z}$ are neglected, as the transverse motion  is quantized, 
so ${\bf B}_{\rm SO}=(\alpha p,\beta p,0)$.} and $\boldsymbol{\sigma }$ are the Pauli spin matrices. The total Hamiltonian, without impurity scattering, also includes the kinetic energy $p^{2}/2m$ and Zeeman 
term $-\mathbf{b}\boldsymbol{\sigma }$, 
where $\mathbf{b}=g\mu _{B}\mathbf{B}/2$. Let us introduce a unit vector $\mathbf{n}$ in the direction $\alpha \hat{x}+\beta \hat{y}$ of ${\bf B}_{\rm SO}$. We also  introduce the
longitudinal and transverse components of magnetic field: $\mathbf{b}=b_{\Vert }\mathbf{n}+\mathbf{b}_{\bot }$. 
The total Hamiltonian reads:
\begin{equation}
H=
p^{2}/2m
+\left( \gamma p-b_{\Vert }\right) \mathbf{n}\boldsymbol{\sigma}
+\mathbf{b}_{\bot }\boldsymbol{\sigma},\quad \gamma =\sqrt{\alpha ^{2}+\beta ^{2}}. \label{Ham}
\end{equation}%
Its eigenvalues are
\begin{equation}
E\left( p,\sigma \right)
=p^{2}/2m
+\sigma q,~q=\sqrt{\left( \gamma
p-b_{\Vert }\right) ^{2}+\mathbf{b}_{\bot }^{2}},  \label{eigenvalues}
\end{equation}%
where $\sigma =\pm 1$ shows the projection of the electron's spin on the total effective Zeeman magnetic field $\mathbf{B}+\mathbf{B}_{\rm SO}$ and is the eigenvalue of the operator $\Sigma =\frac{%
\left\vert \gamma p-b_{\Vert }\right\vert }{q}\left( \mathbf{n}+\frac{%
\mathbf{b}_{\bot }}{\gamma p-b_{\Vert }}\right)\boldsymbol{\sigma} $. For a nonzero
transverse magnetic field $\mathbf{b}_{\perp}$, the direction of spin quantization
depends on momentum. The graph of energy vs. magnetic field for small
magnetic fields $\left\vert \mathbf{b}\right\vert \ll p_{F}^{2}/2m$
is represented by two slightly distorted Rashba parabolas shifted
vertically in opposite directions and with avoided crossing (Fig. 1). The
parallel 
magnetic field $\mathbf{b}_{\Vert}$ is responsible for the reflection asymmetry,
whereas 
$\mathbf{b}_{\perp}$ causes the avoided crossing.  There are generally four Fermi momenta, two left and right movers for each value of $\sigma $.

For a typical experimental setup the SO velocity $\gamma \ll
v_{F}=p_{F}/m$. 
If $\left\vert
\mathbf{b}\right\vert \ll \gamma p_{F}$  then the four
Fermi momenta differ only slightly from the Fermi momentum 
 of  the wire without SO interaction and
magnetic field, $p_{F}=\pi \hslash n/2$ 
($n$ is the 1D electron density), and are given by
\begin{equation}
p_{\sigma \tau }=\tau p_{F}-\sigma m\left[ \gamma -\tau \frac{b_{\Vert }}{%
p_{F}}+\frac{\mathbf{b}_{\bot }^{2}}{{2}p_{F}\left( \gamma p_{F}-\tau
b_{\Vert }\right) }\right] ,  \label{p-sigma}
\end{equation}%
where $\tau =\pm 1$ 
indicates right (R) and left (L) movers.
In the ground state electrons with spin projection $\sigma $
fill a momentum interval from $p_{\sigma -}$ to $p_{\sigma +}$.

All states
in the interval $\left( p_{--},p_{++}\right) $ are  doubly occupied.
A net spin-flip is possible only in the singly occupied intervals
$\left( p_{++},p_{-+}\right) $ and $\left( p_{+-},p_{--}\right)$,
and requires energy $E_{sf}=2q\approx 2\left( \gamma |{p}|-\tau
b_{\Vert }\right) $. Thus, for $\mathbf{b}_{\parallel}\neq 0$, there
are two different resonance frequencies corresponding to the left
and right movers. The \textquotedblleft lengths\textquotedblright\
of
singly-occupied intervals are $2m\left( \gamma -\tau b_{\Vert }/p_{F}%
\right) $. For $m\gamma \ll\hbar n$, we find that the spin-flip
energies are centered at $E_{sf}^{0}=2\left( \gamma p_{F}-\tau
b_{\Vert }\right) $ and confined to narrow bands of width $
4m\gamma \left( \gamma -\tau b_{\Vert }/p_{F}\right) =2m\gamma
E_{sf}^{0}/p_{F}\ll E_{sf}^{0}$. Spin-flip processes can be excited
by a resonant external field with frequency $\omega
_{r}=E_{sf}^{0}/\hslash $. The temperature must be small $T<\hslash
\omega _{r}/k_{B}$ to avoid thermal smearing.

\begin{figure}
\includegraphics[width=2.5in]{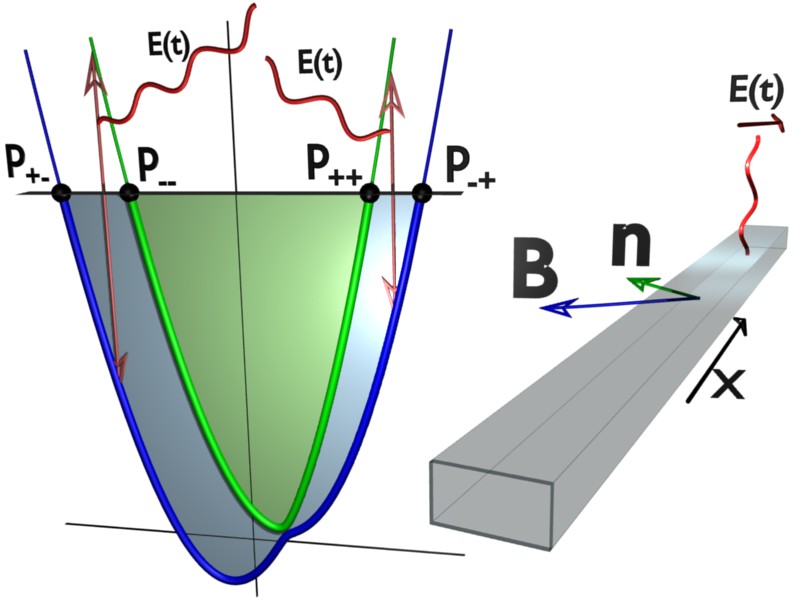} 
\caption{ (color online) Left part: Energy \textit{vs} momentum in given by (\ref{eigenvalues}). 
Thick parts of the spectrum are occupied. The spin reversing  excitations 
of the occupied states by ac electric field are shown. Two
transitions are indicated by long vertical arrows. Right part: The geometry 
and directions of the external magnetic field ${\bf B}$ and internal ${\bf B}_{\rm SO}\parallel \vec{n}$ are shown.} \label{fig:RashbaBands}
\end{figure}

\noindent \textbf{Transition rate by linearly polarized ac field. }
Let an ac electric field
linearly polarized along $x$ direction $\mathbf{E}\left( t\right) =\hat{x}E%
_{0}\left( t\right) e^{-i\omega _{0}t}+\hat{x}E_{0}^{\ast }\left( t\right)
e^{i\omega _{0}t}$ have spectral intensity $I\left( \omega \right) $ centered
about $\omega _{0}$ with width $\Delta \omega \ll \omega _{0}$. Here 
$E_{0}\left( t\right) $ describes a stationary random process with
correlation time $\tau _{B}=\frac{2\pi }{\Delta \omega }$. Averaged over a
time interval $t^{\prime }$ satisfying $2\pi /\omega _{r}\ll t^{\prime }\ll
\tau _{B}$, we have $\overline{E_{0}^{\ast }\left( t\right) E_{0}\left(
t^{\prime }\right) }=\left( 2\pi \right) ^{-1}\int_{-\infty }^{\infty
}I_{\omega }e^{i\omega \left( t-t^{\prime }\right) }d\omega $. It interacts
with the spin since $p$ in the Hamiltonian (\ref{Ham}) must be
replaced by $p+\frac{e}{c}A$ (other directions of polarizations do not couple to spin.) 
In the Weyl gauge (electric
potential $\Phi =0$) 
the relation between the vector-potential $A$ and electric
field reads $A=-\frac{ic}{\omega _{0}}\left( E_{0}\left( t\right)
e^{-i\omega _{0}t}-E_{0}^{\ast }\left( t\right) e^{i\omega _{0}t}\right) $.
Thus, the interaction\ between the electric field and spin is 
$
H_{int}=-\frac{ie\gamma }{\omega _{0}}\left( E_{0}\left( t\right)
e^{-i\omega _{0}t}-E_{0}^{\ast }\left( t\right) e^{i\omega _{0}t}\right)
\mathbf{n}\boldsymbol{\sigma}
$.
For  $\mathbf{b}_{\bot }=0$, the interaction Hamiltonian is proportional to
the same spin projection $\mathbf{n}\boldsymbol{\sigma} $ which enters the static
Hamiltonian (\ref{Ham}) and therefore does not produce spin reversal. Then
only magnetic dipolar transitions can reverse electron's spin. 
However, $\mathbf{b}_{\bot }\ne0$ 
makes electric field induced spin reversal not only possible, but more
probable than magnetic dipolar ones. With the matrix element $\left\langle
+\right\vert \mathbf{n}\boldsymbol{\sigma}\left\vert -\right\rangle =2\left\vert \mathbf{%
b}_{\bot }\right\vert /E_{sf}^{0}$ of the operator 
$\mathbf{n}\boldsymbol{\sigma}$ producing spin reversal between the two eigenstates of the
operator $\Sigma $, time-dependent perturbation theory gives the spin-flip
transition rate for an electron with momentum $p$ as
$ w=\frac{4e^{2}\gamma ^{2}}{\hslash ^{2}\omega
_{0}^{2}}(\mathbf{b}_{\bot }/E_{sf}^{0})^{2}I_{(2\gamma \left\vert
p\right\vert /\hslash )-\omega _{0}} $. In order of magnitude
$I_{\omega }\approx \pi \overline{E_{0}^{2}}/\left( \Delta \omega
\right) $, where $\Delta \omega $ is the spectral width of the ac
field, so
\begin{equation}
w\approx \pi e^{2}\overline{E_{0}^{2}} \left( \mathbf{b}_{\bot
}/E_{sf}^{0}\right) ^{2}/p_{F}^{2}\Delta \omega .  \label{w}
\end{equation}
The ratio of the electric and magnetic transition rates
is $( c\mathbf{b}_{\bot }/v_{F}E_{sf}^{0})^{2}$. The ratio $c/v_{F}$
is about $10^{3}$ for InGaAs. Thus, for a comparatively small
transverse magnetic field $b_{\bot }\sim 10^{-1}E_{sf}^{0}$ the
transition rate (\ref{w}) exceeds magnetic dipolar induced rate 
by 4
orders, whereas the resonance
frequency changes 
only by 1\%.

Perturbation theory is valid if the average occupation number for
excited electrons is small, i.e. $w\teff\ll 1$, where
$\teff$ is a characteristic lifetime. In the ballistic regime
the time of flight $\tau_f=L/v_F$ is much shorter than any collision
time and plays the role of lifetime for an excited electron or hole. In
1D elastic scattering can reverse the excitations' velocity. If
the corresponding back-scattering time $\tau_b$ is much less than
$\tau_f$, then diffusion occurs, with lifetime
$\teff=\tau^2_f/\tau_b\gg\tau_f$. 

If $w\teff\gtrsim 1$,
recombination almost completely compensates the excitation and thus
further increase of power of the external ac field becomes
ineffective. If the spectral width of the external ac field
$\Delta\omega$ is smaller than $\teff^{-1}$, 
then Rabi oscillations occur.

\noindent \textbf{Dynamic generation of dc currents
and magnetization.} If the longitudinal field 
satisfies the
condition $b_{\Vert }\gg {m}{\gamma }^{2}$, the resonance lines for
right and left movers are distinct and can be excited separately.
Thus a resonant linearly polarized ac field can produce a
magnetization as well as dc electric and spin currents. Consider a
linearly polarized ac field that causes spin flips of right movers.
First we obtain upper limits for the stationary density of excited
electrons (and holes) and currents.  

The density of
right-moving states participating in the resonance spin-reversal
process is $n_{sr}=n\Delta\omega/4\omega_r$. As these states have a 
lifetime $\sim \teff $, the density of excitations
can be estimated as
$n_{ex}=\min(w\teff,1)n_{sr}$. The hole
density is the same. The maximum number of excitations
$n_{ex}^{max}=n(\gamma/2v_F)$ occurs for $w\teff\geq 1$,
$\Delta \omega=4m\gamma^2/\hbar$. The pumped spin per
electron is $s=n_{ex}/n$.

An upper limit for the electric current can be obtained on assuming
that all excited electrons move with their initial velocity. The
equilibrium electric current is zero. After excitation an electron
velocity increases by $2\gamma$. Thus, the maximum electric current
is $j_e^{max}=2\gamma en_{ex}^{max}$. The maximum spin current is 
$j_s^{max}=v_Fn_{ex}^{max}$. However, the upper limit usually
is not reached, especially in the diffusive regime where elastic
back-scattering reverses the excitations' velocities. For
$w\teff<1$, in the ballistic regime the electric current is:
\begin{equation}
j_e=2\gamma en_{ex}=enw\tau_f\gamma^2/v_F \label{current-ballistic}
\end{equation}

To show how diffusion affects the currents, for simplicity we
neglect both spin-flip back-scattering and energy relaxation. A
simplified set of kinetic equations reads:
\begin{eqnarray}
&&dn_{R\uparrow }^{e}/dt=w n_{sr}-(\teff^{-1}+
\tau _{b}^{-1})n_{R\uparrow }^{e}+\tau _{b}^{-1}n_{L\uparrow }^{e}, \\
&&dn_{L\uparrow }^{e}/dt=-(\teff^{-1}+\tau _{b}^{-1})
n_{L\uparrow }^{e}+\tau _{b}^{-1}n_{R\uparrow }^{e}.
\label{dynamics}
\end{eqnarray}%
where $n_{R\uparrow }^{e}$ and $n_{L\uparrow }^{e}$ are the
densities of right and left-moving spin up states,
respectively. The ac field creates equal numbers of electrons and
holes with parallel spins, and this property is maintained by the
back-scattering if spin flip process are negligible. The pumped
spin is
polarized approximately along  $\mathbf{n}+\mathbf{b}_{\perp}/\gamma p_{F}$. Its dc absolute value per unit length is $%
s_{eff}=2w\teff n_{sr}$. The spin current is
$j_{s}={g}\mu _{B}v_{F}n_{sr}w\teff\tau_{b}/(2\teff+\tau_{b})$. 
The electric current is:
\begin{equation}
j_{e}=-2ewn_{sr}\frac{\teff\tau _{b}}{2\teff+\tau _{b}}%
\gamma +ewn_{sr}\frac{\teff b_{\perp }^{2}(4\teff+\tau _{b})}{\gamma {p_{F}^{2}}(2\teff+\tau _{b})}.
\label{electric}
\end{equation}%
Eq (\ref{electric}) shows that the electric current changes sign
in diffusive regime at $b_{\perp }>\gamma {p_{F}}\sqrt{\tau
_{b}/2\teff}$. 
%
This happens because the back scattering equalizes the number of left and right moving excitations, whose  velocities differ. 
For resonance of left movers, at frequency $%
\omega _{r}^{L}=2(\gamma {p_{F}}+b_{\Vert })/\hbar $,
the magnetization and currents are reversed.


\noindent \textbf{Relaxation processes}. Relaxation processes play
an extremely important role in spin resonance phenomena. In a typical wire of
10$\mu$m length and 10$\times$10nm$^2$ cross area the number of
electrons is $\sim 1000$. Because this is small, we neglect the
electron-electron interaction. At low temperature the main mechanism
of energy relaxation is Cherenkov emission of phonons. If the
corresponding relaxation time $\tau_{ep}$ becomes comparable to or
shorter than $\tau_f$, energy relaxation occurs before
electrons and holes leave the wire. It does not change the total
spin, but may decrease the excitation's velocity. On the other hand,
energy relaxation removes particles from the excited states and
fills the depleted states. This makes the increase of power of the
external ac field more effective. The electron-phonon interaction is
modeled by a standard Hamiltonian
$H_{ep}=U\int\nabla\bf{u}(\bf{x})\psi^{\dagger}(\bf{x})\psi(\bf{x})$,
where $\bf{u}(\bf{x})$ is the displacement vector, $\psi(\bf{x})$ is
the
 electron 
field operator and $U$ is the deformation potential.
Electrons in the wire are always one-dimensional, but phonons can be
1,2, or 3-dimensional depending on the experimental setup. For
example, for an electron with momentum deviating by $\xi$ from the
Fermi-point, and emitting 3D phonons, a slightly simplified result
for the relaxation time is $\tau_{ep}^{-1} =
\frac{U^2a^3}{6\pi\hslash Muv_F}(v_F\xi/\hslash u)^3$,
where $M$ and $a^3$ are the mass and lattice constant of elementary
cell;  
$u$ is the sound velocity.

%
In 2D and 3D systems elastic scattering (diffusion) leads to spin
relaxation  by the Dyakonov-Perel mechanism \cite{dyakonov-perel,*dyakonov-perel1}
because the direction of the internal Zeeman field ${\bf B}_{\rm SO}$ depends on the
direction of the electron's momentum and is randomized by diffusion.
In 1D for $\mathbf{b}_{\perp }=0$ the direction of  ${\bf B}_{\rm SO}$ 
is the same for all electrons and Dyakonov-Perel mechanism does not work.
For $\mathbf{b}_{\perp}\neq 0$, spin flip does occur  in back scattering. 
However, 
 in the resonance regime $\omega_r\tau_b\gg 1$ 
the direction of ${\bf B}_{\rm SO}$  is not random during an oscillation. 
Spin-flip then has a probability  of the order of
$(\mathbf{b}_{\perp}/E_{sf}^0)^2$ and can be
neglected. Other spin relaxation mechanisms, such
as phonon emission combined with SO interaction, are much weaker.

\noindent \textbf{Numerical estimates}. All numerical estimates are
made for In$_{0.53}$Ga$_{0.47}$As. We take $m=4.3\times 10^{-29}$g,
$\alpha =1.08\times 10^{6}$cm/s, and $g=-0.5$.\cite{[{For
simplicity we neglect the anisotropy of the g-factor in a wire.
Experimentally it is not small, }][{, and must be taken into account in a
more accurate theory.}]dorokhin,note} A typical 2D electron density
is $2\times 10^{12}$cm$^{-2}$. Let a wire width be
$a=5$nm and length $l=$ 1-10 $\mu $m. Then we find 1D
density $n=10^{6}$cm$^{-1}$, 
$p_{F}=1.65\times 10^{-21}$g-cm/s and 
$v_{F}=0.38\times 10^{8}$cm/s. Assuming $\alpha =\beta $, we have
$\omega _{r}=4.8\times 10^{12}$s$^{-1}$ ($\sim 0.8$THz) and the resonance 
width $\Delta \omega$ is $3.8\times 10^{11}$s$%
^{-1}$. The value $\overline{E_{x}^{2}}$  in Eq. (\ref{w}) is
determined by the source power in the terahertz range. Although
standard cascade lasers have power in the range 1mW -
1W,\cite{tera,*tera1} the power can be strongly enhanced by
non-linear devices, and in very short pulses (1 ps) it can reach
1MW.\cite{highfield,highfield1,highfield2} The free-electron laser
at UCSB provides continuous power 1-6 kW  for the frequency range
0.9-4.75THz. On focusing, the energy flux becomes up to $40
\textrm{kW}/\textrm{cm}^2$.\cite{*[{}][{ Also see the website
\url{http://sbfel3.ucsb.edu}}]kono}
For the moderate flux $S=1$kW/cm$^2$ , we find $\overline{E^{2}}=S/c 
=0.33$erg$\cdot$cm$^{-3}$. 
 For $B_{\bot }=10$T we have $b_{\bot }/E_{sf}^0=0.048$, and Eq. (\ref{w})
yields $w=0.48\times 10^{9}$s$^{-1}$. As noted above, $w$ can be
increased by changing the power or size of the focus area. For the
length 1-10 $\mu$m the time of flight is $\tau _{f}=2.6\times (
10^{-12}$-$10^{-11}) $s. The back-scattering time $\tau _{b}$ can
be
estimated from typical mobilities 
$\mu =e\tau /m= (2\times10^{4}-4\times10^{5})$cm$^{2}$/Vs
 in the bulk or film.\cite{yamada} Since the scattering
cross-section area is much less than the wire cross-section area,
$\tau $ can be identified with $\tau_{b}$. Typical values are $\tau
_{b}=2.7\times (10^{-13}$-$10^{-12}) $s. In this case the regime is
either diffusive or marginal between diffusive and ballistic. In the
ballistic regime with $\tau _{f}=2.6\times 10^{-12}$s, according to
Eq. (\ref{current-ballistic}) the electric current equals 
$18$nA.
The spin current is 
$j_{s}=8.3\times 10^{9}$s$^{-1}$.
In the diffusive regime $\left( \tau_{b}=2.7\times 10^{-12}\,{\rm
s},\tau _{f}=2.6\times 10^{-11}\,{\rm s}\right) $, according to
Eq.(\ref{electric}) the electric current $I_{e}$ in the wire
equals  $3.6$ pA and the magnetization per particle, in Bohr
magnetons, is $\frac{n_{ex}}{n}w\teff\sim 0.002$. The
temperature must be maintained below $2\gamma p_{F}/k_{B}\approx
35K$. $w\teff $ is about 0.12, indicating that perturbation
theory works well and there is a possibility for current and spin
increase.

For energy relaxation we assume 3D phonons and take numerical
values for InGaAs: deformation potential $U=16$eV, sound velocity
$u=3.3\times 10^{5}$cm/s,\cite{sugaya} $a=5$\AA, 
$M=1.8\times 10^{-22}$g and $\xi =m\gamma$. Then the energy
relaxation time $\tau_{ep}=1.4\times 10^{-12}$s. With 2D and 1D
phonons the formulae differ, but numerical estimates give the same
order of magnitude. This result shows that, even  in the ballistic
regime, $\tau_{ep}$ is usually much shorter than $\tau_f$ and energy
relaxation is substantial, decreasing the currents.

\noindent \textbf{Prior work}. The generation of currents by an ac field is similar to the photogalvanic
effect predicted by Ivchenko amd Pikus \cite{ivchenko-pikus}
and by Belinicher \cite{belinicher}. More recently many clever modifications
of this effect have been proposed and experimentally observed (see review
\cite{ganichev}, and article \cite{ivchenko,*ivchenko2}).
They are mostly realized in 2D
systems, but more importantly, unlike our case \textit{non-resonance} optical or infrared radiation is used.
The resonant nature of our effect also ensures
the strong dependence of the resonance line
and transition probability on the transverse magnetic field.

Shechter
{\textit{e}t al.}
considered a similar problem in a 2D film with SO interaction.
\cite{finkel} They noted that Rashba SO  \cite{rashba,*BychkovRashba}
leads to what they called a \textit{chiral resonance.}
Transitions between states with different chirality  
are possible only for electrons with momenta
located between two Fermi-circles corresponding to different chiralities.
However, as the authors themselves noted,
the Dresselhaus SO interaction \cite{dressel} strongly broadens the
resonance.
This broadening is of the same order of magnitude as the spin-flip energy. In Ref.~%
\onlinecite{finkel} the authors proposed to avoid this broadening by
choosing a specific growth direction, or by decreasing the electron density.
Both proposals require sophisticated experimental techniques.

\noindent \textbf{Conclusions}. We predict that in 1D the SO interaction gives
rise to spin resonance, even without an external magnetic field. This
happens because for a 1D degenerate Fermi gas the direction of the
effective SO magnetic field is the same for all $\mathbf{p}$. The
resonance frequency is typically in the terahertz region with
relative width depending linearly on the Dresselhaus and Rashba SO
constants. The external longitudinal magnetic field (parallel to the
internal SO field) separates the resonance frequencies of the left
and right movers, producing charge and spin currents. A
perpendicular magnetic field violates reflection symmetry and
couples an ac electric field to the spin which otherwise is flipped
only by
the weak magnetic dipolar interaction. 
On resonance an ac electric field linearly polarized along the wire
produces dc charge current, dc spin current, and dc magnetization. The
amplitude of these effects can be easily controlled by the static external
magnetic field and the gate voltage.

\noindent \textbf{Acknowledgements}. We thank A.M. Finkelstein and
M. Khodas for discussions and for pointing out their
work,\cite{finkel} and J. Kono for an illuminating discussion of the
experimental situation. This work was supported by the Department of
Energy under the grant DE-FG02-06ER46278. Ar.A. was supported by
NSF0757992 and Welch Foundation (A-1678).
\bibliography{spin-resonance}

\end{document}